\documentclass[a4paper,11pt]{article}
\usepackage{pos}
\graphicspath{{Figures/}}
\title{ Infrared Cloud Monitoring with UCIRC2}

\ShortTitle{UCIRC2}

\author*[a]{Rebecca Diesing}
\author[a]{Stephan S. Meyer}
\author[a]{Johannes Eser}
\author[a]{Alexa Bukowski} 
\author[a]{Alex Miller}
\author[a]{Jake Apfel}
\author[a]{Gerard Beck}
\author[a]{Angela V. Olinto}

\affiliation[a]{Department of Astronomy \& Astrophysics, KICP, EFI, The University of Chicago, IL 60637, USA\\}

\onbehalf{for the JEM-EUSO collaboration} 

\emailAdd{rrdiesing@uchicago.edu}

\abstract{The second generation of the Extreme Universe Space Observatory on a Super Pressure Balloon (EUSO-SPB2) is a balloon instrument that searched for ultra high energy cosmic rays (UHECRs) with energies above 1 EeV and very high energy neutrinos with energies above 1 PeV. EUSO-SPB2 consists of two telescopes: a fluorescence telescope pointed downward for the detection of UHECRs and a Cherenkov telescope toward the limb for the detection of PeV-scale showers produced by neutrino-sourced tau decay (just below the limb) and by cosmic rays (just above the limb). Clouds inside the fields of view of these telescopes—particularly that of the fluorescence telescope—reduce EUSO-SPB2's geometric aperture. As such, cloud coverage and cloud-top altitude within the field of view of the fluorescence telescope must be monitored throughout data-taking. The University of Chicago Infrared Camera (UCIRC2) monitored these clouds using two infrared cameras centered at 10 and 12 $\mu$m. By capturing images at wavelengths spanning the cloud thermal emission peak, UCIRC2 measured cloud color-temperatures and thus cloud-top altitudes. In this contribution, we provide an overview of UCIRC2, including an update on its construction and performance. We also show first results from the flight.}

\ConferenceLogo{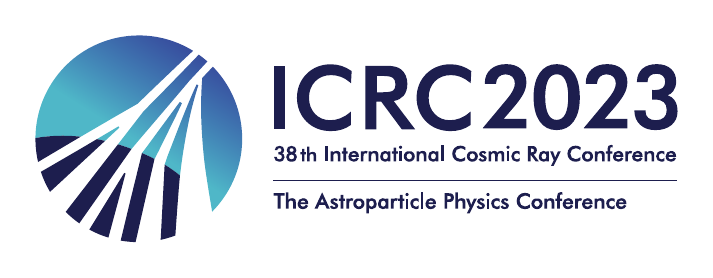}

\FullConference{%
38th International Cosmic Ray Conference (ICRC2023)\\
  26 July - 3 August, 2023\\
  Nagoya, Japan}

\begin{document}
\maketitle

\section{Introduction}

Ultra High Energy Cosmic Rays (UHECRs), cosmic rays (CRs) with energies above $10^{18}$ eV, are currently detected with ground-based observatories such as the Telescope Array \cite{TA12} in Utah and the Pierre Auger Observatory \cite{auger15} in Argentina. In particular, UHECRs can be detected via the characteristic particle shower, called an Extensive Air-Shower (EAS), that occurs when an UHECR interacts with Earth's atmosphere. This EAS produces fluorescence of atmospheric nitrogen molecules, detectable in the 300-400 nm spectral band, as well as optical Cherenkov light. Because UHECRs are rare, ($<$ 1 per km${^2}$ per century close to $10^{20}$ eV), charged-particle astronomy requires extremely large detector volumes. One way to increase detector volume is to observe the atmosphere from above. This technique was tested by the Extreme Universe Space Observatory on a Super Pressure Balloon (EUSO-SPB2) during a brief flight in the spring of 2023.

\begin{figure}
\begin{center}
\includegraphics[width=0.95\textwidth, clip=true,trim= 0 0 20 0]{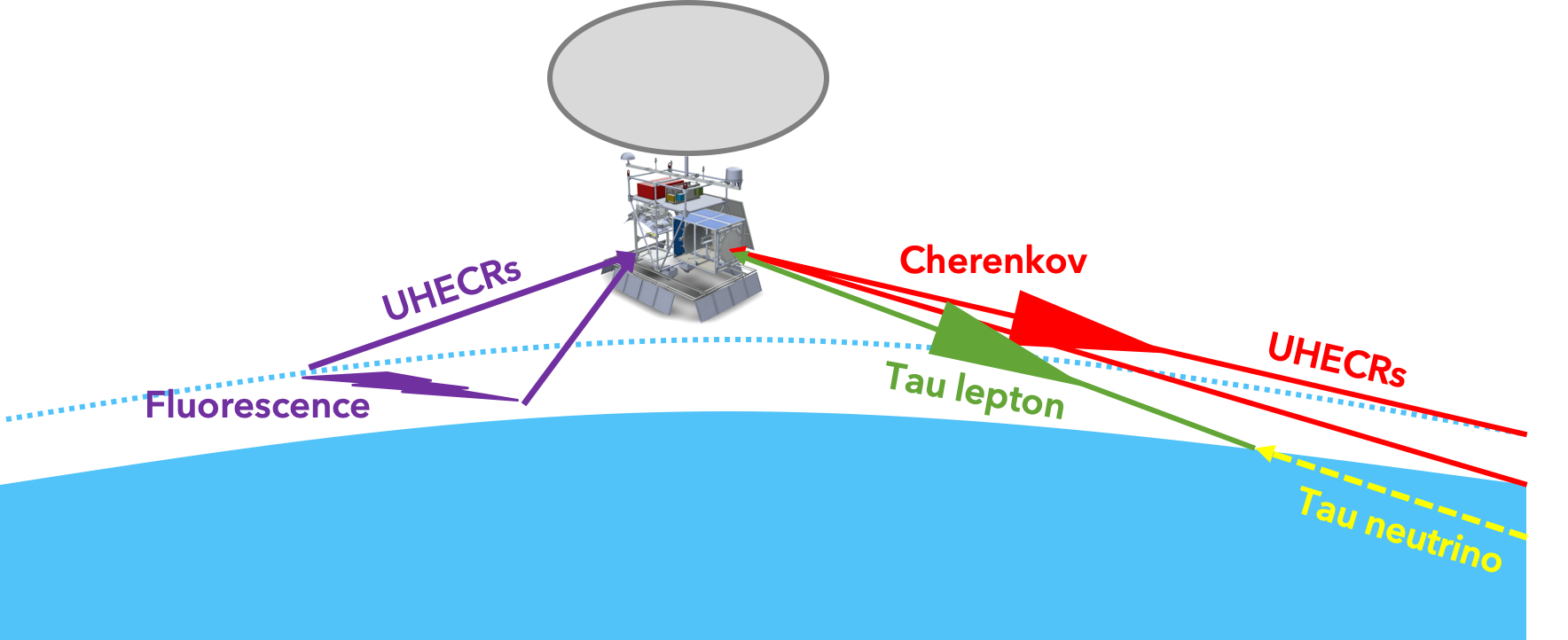}   
\end{center}
\caption{EUSO-SPB2's three detection modes: fluorescence from UHECRs (purple), Cherenkov from UHECRs (red), and Cherenkov from CNs (green).}
\label{fig:EUSOSPB2}
\end{figure}

A pathfinder to a more ambitious sattelite mission, EUSO-SPB2 can detect UHECRs via two complementary techniques: looking down upon the atmosphere with a fluorescence telescope and looking towards the limb of the Earth to observe the Cherenkov signals produced by UHECRs above the limb. EUSO-SPB2 also searched for the signatures of neutrinos above $10^{16}$ eV via the Cherenkov light from upward going tau leptons produced when a tau neutrino interacts near the surface of the Earth (see Figure \ref{fig:EUSOSPB2}) \cite{SPB17}. 

The presence of high clouds within the detectors' field of view (FoV)--particularly the fluorescence telescope--can significantly reduce the UHECR event detection rate and event energy calibration. Namely, it is possible for some of the EAS signal to occur behind high clouds. Determining the EUSO-SPB2s exposure to UHECRs thus requires knowledge of the effective detector volume, i.e., the volume of atmosphere within the FoV, above the clouds. Thus, EUSO-SPB2 requires continuous information about cloud coverage and altitude. This is the responsibility of the second generation of the University of Chicago Infrared Camera (UCIRC2). In this proceeding, we present an overview of the UCIRC2 instrument, including preliminary results from its 2023 flight.

\section{Method}

\begin{figure}[ht]
    \centering
    \includegraphics[width=0.95\textwidth]{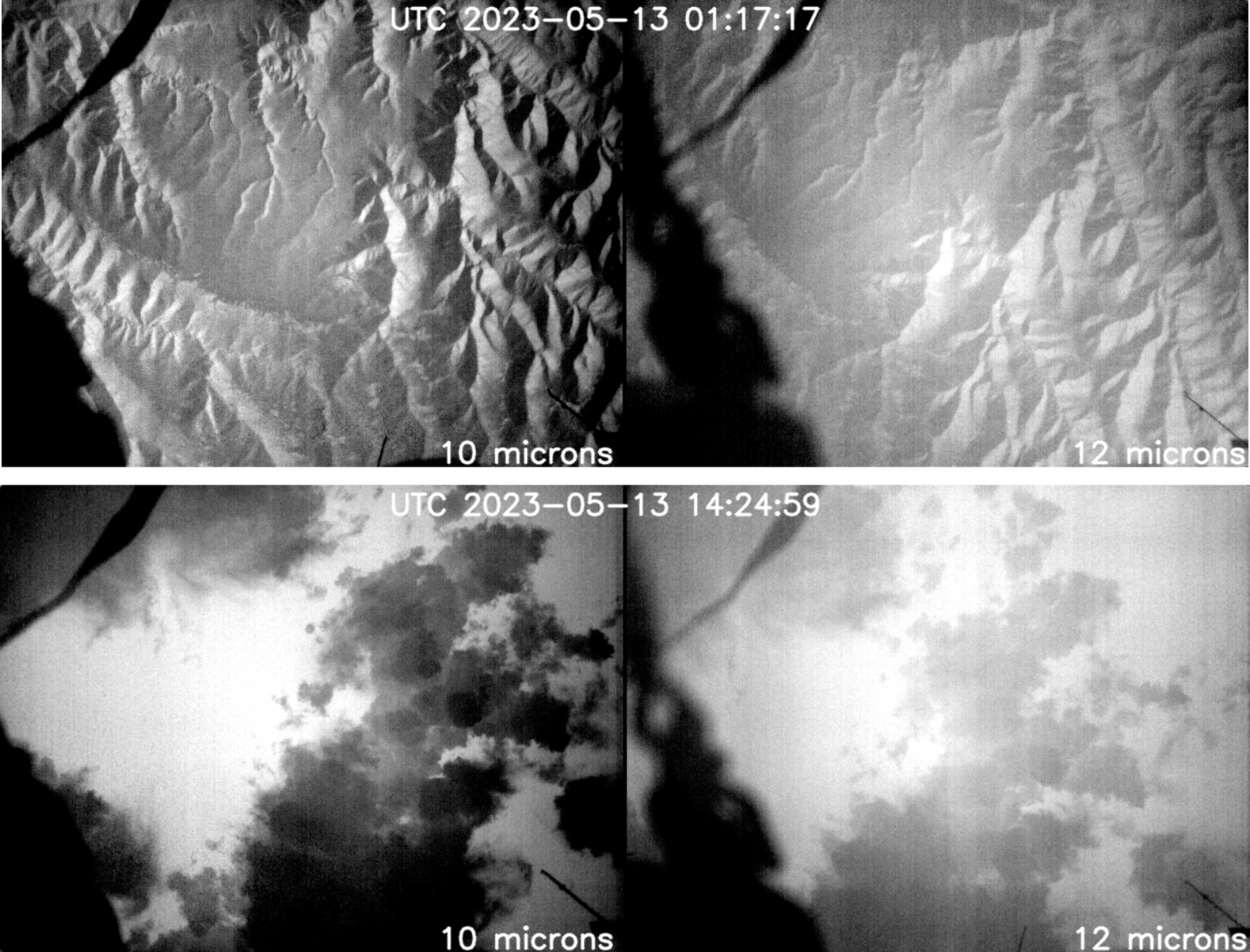}
    \caption{Uncalibrated images of mountains (top) during the daytime and clouds (bottom) during the night captured by UCIRC2, which flew on EUSO-SPB2 in the spring of 2023. Also visible in the foreground is a portion of the EUSO-SPB2 gondola (lower left corners of each image) as well as cables and antennas hanging from the gondola. Using an IR camera, cloud coverage can be easily determined. Cloud temperature (and thus altitude) can be determined by observing at two wavelengths near the cloud blackbody peak, in our case 10 and 12 $\mu$m.}
    \label{fig:SamplePic}
\end{figure}

When EUSO-SPB2 is in observing (night) mode, IR images of the environmental conditions in and around the effective UHECR detection area are captured by UCIRC2 every 120 seconds. These images can be used to collect information about cloud coverage and altitude (cloud top height, CTH) within the field of view of the UHECR detectors (see Figure \ref{fig:SamplePic}).

Because the clouds are at the temperature of the air, CTH can be inferred from cloud temperature, $T_{\rm c}$ which can be estimated using two brightness temperatures in bands near in wavelength to the cloud blackbody peak. More specifically, UCIRC2's two IR cameras observe at wavelengths of 10$\mu$m and one at 12$\mu$m. A calibrated image in a single frequency band can be used to determine the temperature of an object of known emissivity ($\epsilon$), but cloud emissivity is highly variable and significantly less than 1. Thus, a multifrequency observation is required to break the degeneracy between $\epsilon$ and $T_{\rm c}$. For a single layer of clouds above an ocean of known surface temperature and reflectivity (and thus power, $P_{\rm E}$), one can estimate power on the detector, $P_{\rm tot}$ as,

\begin{equation}
    P_{\rm tot} = \epsilon P_{c}+(1-\epsilon)P_{E}.
\end{equation}
Here, $P_{c}$ is the power of the cloud, from which $T_{c}$ and thus CTH, can be inferred. Other methods for reconstructing CTH can be found in \cite{anzalone+19}.

\begin{figure}
    \centering
    \includegraphics[width=\textwidth]{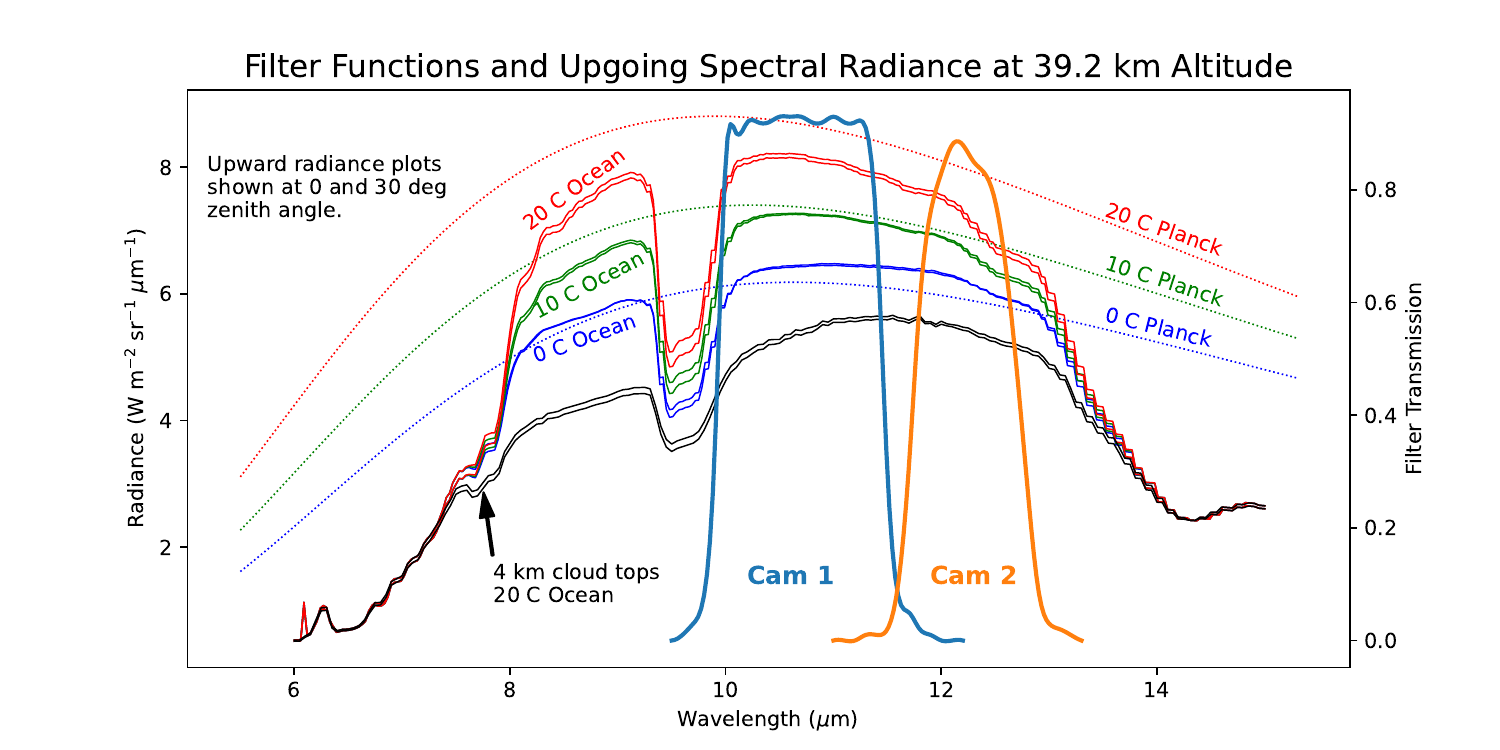}
    \caption{Upgoing spectral radiance as a function of wavelength as calculated using COART \cite{jin+06}, assuming different ocean temperatures with no clouds (red, green, and blue lines) and the presence of clouds with tops 4 km above a 20 C ocean (black lines). The two solid lines shown in each case correspond to 0 and 30 degree zenith angles. The blackbody curves corresponding to each ocean temperature are also shown for reference (dotted lines), as are the bandpasses of UCIRC2's two filters. We will use this model to determine the altitudes of clouds beneath EUSO-SPB2's fluorescence detector.}
    \label{fig:COART}
\end{figure}

For a more precise calculation, we use the Coupled Ocean-Atmosphere Radiative Transfer Model (COART) presented in \cite{jin+06}, which calculates the radiance at any frequency by solving the radiative transfer equation from the ocean to a specified level in the atmosphere, including clouds of arbitrary altitude and emissivity (see Figure \ref{fig:COART}).

\section{Design}

\paragraph{IR Cameras.}

UCIRC2 is outfitted with two $640\times480$ pixel Teledyne DALSA Calibir GXF uncooled IR cameras with 14mm lenses, focused at infinity. The cameras have a $42^\circ \times 32^\circ$ FoV, chosen to be somewhat larger than that of EUSO-SPB2's fluorescence telescope. When the payload is in ``night mode'', which occurs when the atmosphere is dark enough to allow for proper functioning of photodetection modules (PDMs), UCIRC takes a pair of pictures every two minutes. The wide field of view of the IR cameras  makes it possible to extrapolate the cloud conditions in the section of the atmosphere swept out by the PDM field of view in the time between pictures. 

The native spectral response of the cameras is 8 to 14 $\mu$m, but each camera is fitted with a filter to facilitate the radiative CTH reconstruction. One of the cameras is fitted with an Edmund Optics bandpass light filter that transmits wavelengths between 9.6 and 11.6$\mu$m (denoted 10$\mu$m); the other is fitted with a SPECTROGON bandpass light filter which transmits wavelengths between 11.5 and 12.9$\mu$m (denoted 12$\mu$m). These bands are spaced to obtain brightness temperature data that facilitates both the Blackbody Power Ratio CTH reconstruction and the Radiative Transfer Equation CTH reconstruction methods discussed in the preceding section. 

The cameras are powered via a 12V connection and communicate via Gigabit Ethernet with a single-board, industrial-grade CPU that can operate at temperatures between -40C and 85C. 

\paragraph{Environment Control.}

\begin{figure}
    \centering
    \includegraphics[width = 0.8\textwidth]{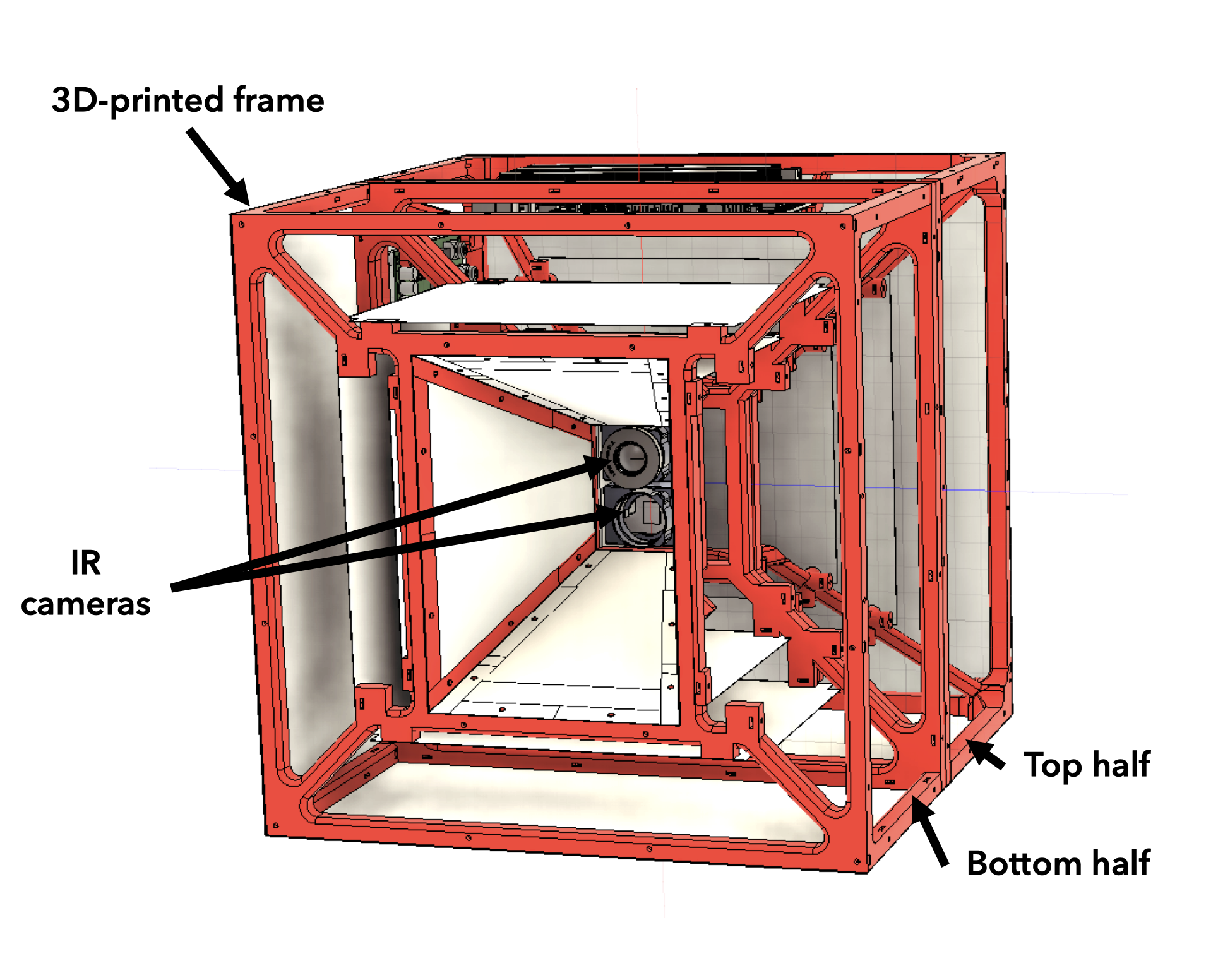}
    \caption{Drawing of UCIRC2, including its 3D-printed frame and  two cameras located near the center of the box (pointed toward the viewer). To show the interior structure of the box, this drawing does not include the painted aluminum panels mounted to UCIRC2's sides.}
    \label{fig:schematic}
\end{figure}

UCIRC2 is designed to operate in a high altitude ($\approx 33$km) environment during both daytime, when ambient temperatures reaches approximately 40C, and nighttime, when ambient temperatures reach approximately -40C. Temperature management is therefore a central design concern. In particular, the camera response is temperature dependent, meaning that camera temperature must be held approximately constant during operation (night mode). To maintain a stable temperature, the two cameras are housed in a $300$mm$\times$300mm$\times300$mm aluminum box coated with high emissivity flat white paint. This box splits into two halves to allow easy access to the cameras and electronics (see Figure \ref{fig:schematic}). 

A temperature management system consisting of resistive heaters and thermometers enables precise temperature monitoring and control. This heating system is controlled by a Meerstetter Engineering HV-1123 thermoelectric cooling and heating controller (TEC). Note that system is designed to be most effective at heating because, in general, UCIRC2 collects data during the nighttime, when the environment is cold. The temperature regulated camera stage is a machined aluminum plate to which both IR cameras are thermally coupled.
Note that the set point temperature for the cameras can be modified by telemetery command, with daytime and nighttime operating temperatures chosen to minimize power consumption. 

\section{Testing and Calibration}

To replicate the expected flight environment, UCIRC2 was tested in a thermovac chamber pumped down to 0.3 mbar and a shroud cooled with liquid nitrogen vapor. The temperature management system was tested over all possible environmental temperatures to ensure that the cameras can be maintained within their operating temperature range. 

\begin{figure}[ht]
    \centering
    \includegraphics[width=\textwidth, clip=true,trim= 0 400 0 0]{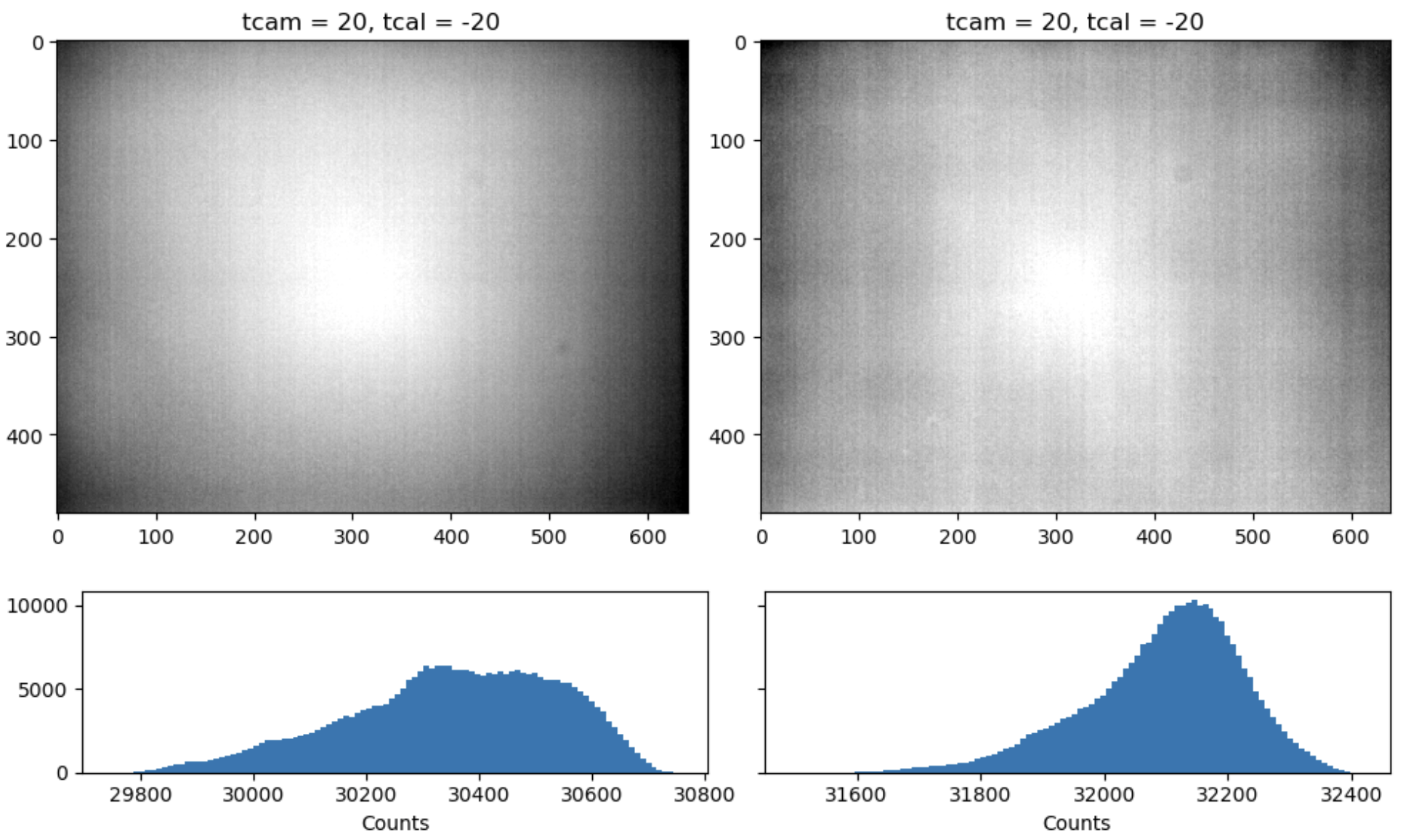}
    \caption{Sample calibration (flat-field) image from each camera, with the camera centered at 10 $\mu$m (12 $\mu$m) on the left (right), and both the cameras and calibration target set to 20C. The resulting pattern is a camera-specific additive offset, which is subtracted from the images taken during flight.}
    \label{fig:calibration}
\end{figure}

To calibrate the cameras, UCIRC2 was then positioned above a calibration target consisting of a highly emissive, temperature-controlled material. By taking of images of the calibration target at multiple temperatures, this target can be used to perform a pixel by pixel calibration of each camera (see Figure \ref{fig:calibration}). Because the cameras' thermal response depends on their temperature, calibration images were also taken at multiple camera temperatures.

\section{Preliminary Results}

\paragraph{Flight Overview.}

\begin{figure}[ht]
    \centering
    \includegraphics[width = \textwidth]{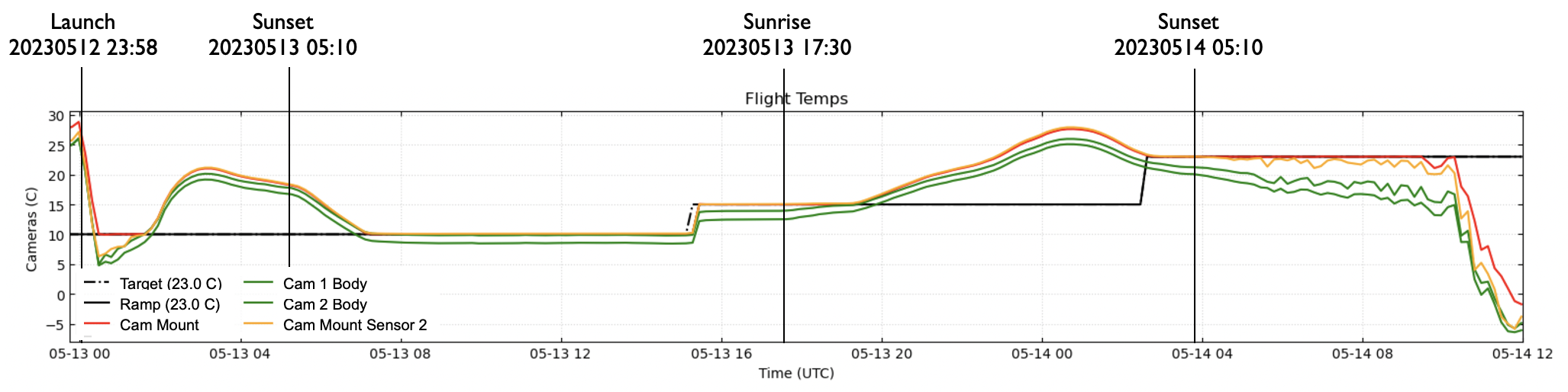}
    \caption{UCIRC2 camera temperatures during flight as a function of time. During nighttime operations, the cameras were maintained at a constant temperature, with control lost only during EUSO-SPB2's descent.}
    \label{fig:temperature}
\end{figure}

During EUSO-SPB2's brief ($\lesssim$ 2 day) flight, UCIRC2 was able to capture one full night's worth of cloud data (i.e., with images captured every two minutes), in conjunction with observations taken by EUSO-SPB2's fluorescence telescope. All of these data were successfully telemetered to the ground. During nighttime operations, UCIRC2 maintained a constant temperature of 10 and later 15C (see Figure \ref{fig:temperature}).

\paragraph{In-Flight Calibration.}

\begin{figure}[ht]
    \centering
    \includegraphics[width = \textwidth]{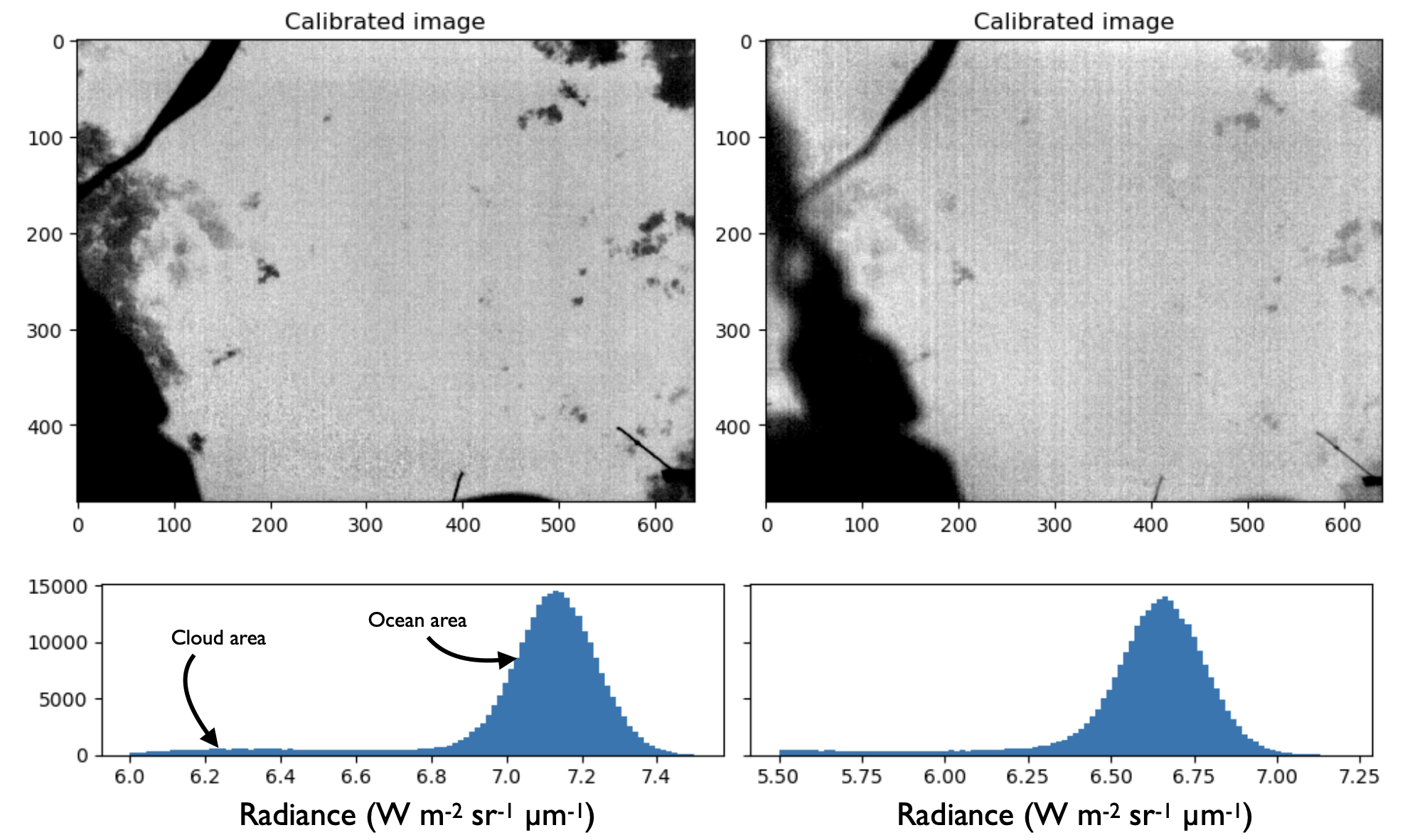}
    \caption{A sample image from each camera (top panels) after applying the calibration technique described in the text. Histograms of the pixels in these images (bottom panels) clearly distinguish between the radiance of the ocean and that of the foreground (clouds, gondola components).}
    \label{fig:CalPic}
\end{figure}

In addition to the thermovac tests described in the preceding section, we performed in-flight calibration by using the ocean as a flat field. Our method proceeds as follows:

\begin{enumerate}
    \item Choose a relatively cloud-free image in which the ocean is visible in most pixels.
    \item Subtract a flat-field image ($I_{\rm ff}$) taken in the thermovac (e.g., Figure \ref{fig:calibration}). Use the same calibration data to estimate camera responsivity (counts per unit of radiance) in each pixel ($I_{\rm r}$).
    \item Using a mask, remove pixels that correspond to foreground objects.
    \item Fit the remaining, flat-field subtracted image to a second-order surface, $I_{\rm ocean}$ (i.e., use the ocean as an additional calibrator). 
\end{enumerate}
To then calibrate an arbitrary image, $I_{\rm init}$ (i.e., to measure the radiance impinging on in each pixel), we estimate,

\begin{equation}
    I_{\rm cal} = \frac{I_{\rm init} - I_{\rm ff} - I_{\rm ocean}}{I_{\rm r}} + I_{\rm COART},
\end{equation}
where $I_{\rm COART}$ is the radiance from a 10 C ocean as a function of zenith angle, as predicted by the COART model described previously. A sample calibrated image is shown in Figure \ref{fig:CalPic}.

\paragraph{Outlook.}

Our calibrated images clearly show cloud coverage, as well as variations in cloud temperature. Going forward, we will use the COART model to make detailed estimates of CTH in order to constrain EUSO-SPB2's aperture during flight. This exercise will also inform design decisions for IR cameras on future missions to detect EAS's from above.

\section{Acknowledgements}
The authors acknowledge the support by NASA awards 11-APRA-0058, 16-APROBES16-0023, 17-APRA17-0066, NNX17AJ82G, NNX13AH54G, 80NSSC18K0246, 80NSSC18K0473, 80NSSC19K0626, 80NSSC18K0464, 80NSSC22K1488, 80NSSC19K0627 and 80NSSC22K0426, the French space agency CNES, National Science Centre in Poland grant n. 2017/27/B/ST9/02162, and by ASI-INFN agreement n. 2021-8-HH.0 and its amendments. This research used resources of the US National Energy Research Scientific Computing Center (NERSC), the DOE Science User Facility operated under Contract No. DE-AC02-05CH11231. We acknowledge the NASA BPO and CSBF staffs for their extensive support. We also acknowledge the invaluable contributions of the administrative and technical staffs at our home institutions.

\bibliographystyle{aasjournal}

\clearpage
\newpage
{\Large\bf Full Authors list: The JEM-EUSO Collaboration\\}

\begin{sloppypar}
{\small \noindent
S.~Abe$^{ff}$, 
J.H.~Adams Jr.$^{ld}$, 
D.~Allard$^{cb}$,
P.~Alldredge$^{ld}$,
R.~Aloisio$^{ep}$,
L.~Anchordoqui$^{le}$,
A.~Anzalone$^{ed,eh}$, 
E.~Arnone$^{ek,el}$,
M.~Bagheri$^{lh}$,
B.~Baret$^{cb}$
% ...
}
\end{sloppypar}
\vspace*{.3cm}

{ \footnotesize
\noindent
% ...
$^{cb}$ Universit\'e de Paris, CNRS, AstroParticule et Cosmologie, F-75013 Paris, France\\
% ...
$^{ed}$ Istituto Nazionale di Fisica Nucleare - Sezione di Catania, Italy\\
% ...
$^{eh}$ INAF - Istituto di Astrofisica Spaziale e Fisica Cosmica di Palermo, Italy\\
% ...
$^{ek}$ Istituto Nazionale di Fisica Nucleare - Sezione di Torino, Italy\\
$^{el}$ Dipartimento di Fisica, Universit\`a di Torino, Italy\\
% ...
$^{ep}$ Gran Sasso Science Institute, L'Aquila, Italy\\
% ...
$^{ff}$ Nihon University Chiyoda, Tokyo, Japan\\ 
% ...
$^{ld}$ University of Alabama in Huntsville, Huntsville, AL, USA\\
$^{le}$ Lehman College, City University of New York (CUNY), NY, USA\\
% ...
$^{lh}$ Georgia Institute of Technology, USA\\
% ...
}

\end{document}